\documentclass[%
 reprint,superscriptaddress,
 amsmath,amssymb,twocolumn
]{revtex4-1}

\usepackage[utf8]{inputenc}
\usepackage{graphicx}
\usepackage{subcaption}
\usepackage{siunitx}
\usepackage{xfrac}
\usepackage{xcolor}
\usepackage{array}
\usepackage{url}
\usepackage{placeins}
\usepackage{natbib}
\graphicspath{{figures/}}

\begin{document}

\title{Identifying observable carrier-envelope phase effects in Laser Wakefield Acceleration with near-single-cycle pulses}
\date{\today}

\author{Julius Huijts}
 \affiliation{LOA, CNRS, Ecole Polytechnique, ENSTA Paris, Institut Polytechnique de Paris, 181 Chemin de la Hunière et des Joncherettes, 91120 Palaiseau, France}
 
\author{Igor A. Andriyash}
 \affiliation{LOA, CNRS, Ecole Polytechnique, ENSTA Paris, Institut Polytechnique de Paris, 181 Chemin de la Hunière et des Joncherettes, 91120 Palaiseau, France}
 
\author{Lucas Rovige}
 \affiliation{LOA, CNRS, Ecole Polytechnique, ENSTA Paris, Institut Polytechnique de Paris, 181 Chemin de la Hunière et des Joncherettes, 91120 Palaiseau, France}
 
\author{Aline Vernier}
 \affiliation{LOA, CNRS, Ecole Polytechnique, ENSTA Paris, Institut Polytechnique de Paris, 181 Chemin de la Hunière et des Joncherettes, 91120 Palaiseau, France}
 
\author{Jérôme Faure}
 \affiliation{LOA, CNRS, Ecole Polytechnique, ENSTA Paris, Institut Polytechnique de Paris, 181 Chemin de la Hunière et des Joncherettes, 91120 Palaiseau, France}

\begin{abstract}
Driving laser wakefield acceleration with extremely short, near single-cycle laser pulses is crucial to the realisation of an electron source that can operate at kHz-repetition rate while relying on modest laser energy. It is also interesting from a fundamental point of view, as the ponderomotive approximation is no longer valid for such short pulses. Through particle-in-cell simulations, we show how the plasma response becomes asymmetric in the plane of laser polarization, and dependent on the carrier-envelope phase (CEP) of the laser pulse. For the case of self-injection, this in turn strongly affects the initial conditions of injected electrons, causing collective betatron oscillations of the electron beam. As a result, the electron beam pointing, electron energy spectrum  and the direction of emitted betatron radiation become CEP-dependent. For injection in a density gradient the effect on beam pointing is reduced and the electron energy spectrum is CEP-independent, as electron injection is mostly longitudinal and mainly determined by the density gradient. Our results highlight the importance of controlling the CEP in this regime for producing stable and reproducible relativistic electron beams and identify how CEP effects may be observed in experiments. In the future, CEP control may become an additional tool to control the energy spectrum or pointing of the accelerated electron beam. 
\end{abstract}

\maketitle

\section{Introduction}
Laser wakefield accelerators (LWFA) \citep{tajima_laser_1979,esarey_physics_2009,faure_laserplasma_2004, mangles_monoenergetic_2004,geddes_high-quality_2004} are table-top setups that can produce relativistic electron bunches with an extremely short (fs) bunch duration \citep{lundh_few_2011} and a small (\si{\um}) source size \citep{ta_phuoc_analysis_2008}. These accelerators are based on an intense laser pulse which interacts with a plasma, exciting a plasma wave which accelerates electrons through extremely high electric fields (in excess of 100 \si{\GeV~\m^{-1}}). As these setups are laser-based, the synchronisation between the electron bunch and the laser pulse is jitter free, making this source ideal for pump-probe experiments requiring high temporal resolution, with applications in solid state physics, materials science and biology \citep{he_high_2013, mahieu_probing_2018, malka_principles_2008,behm_demonstration_2020}. 

Results obtained so far in the field of laser wakefield acceleration have mostly been on systems with a repetition rate on the order of a hertz. Part of the community is currently pushing laser plasma accelerators to a high repetition rate (kHz) \citep{faure_review_2018,salehi_mev_2017}, which is beneficial for applications requiring statistical averaging. In addition, increasing the repetition rate also improves the source stability by reducing thermal variations in the laser system, and enables active feedback control. Increasing the repetition rate to a kHz-level with current Ti:Sapph laser technology means restricting the pulse energy below the 100 mJ-level. The high intensities necessary for LWFA ($10^{18}-10^{19}$ W~cm$^{-2}$) can then still be achieved by compressing the pulse down to the (almost) single-cycle regime, as has recently been shown in \citep{guenot_relativistic_2017,ouille_relativistic-intensity_2020}.

In this regime, the ponderomotive approximation that is generally used in the field of laser-plasma interaction is no longer valid: the pulse envelope varies considerably over an oscillation of the electric field. The dynamics can therefore not be properly described with the pulse intensity (as is the case in the ponderomotive approximation), but the electric field itself needs to be considered instead. In this context, the carrier-envelope phase starts to become relevant, both from a theoretical point of view and as an experimental parameter that needs to be controlled in order to obtain a stable, reproducible accelerator. With the continuing progress of laser technology, CEP control might become available not only for stabilizing, but also for tuning the beam parameters, in ways similar to what is achieved in the field of high harmonic generation \citep{baltuska_attosecond_2003, ishii_carrier-envelope_2014}. In addition, CEP effects can also become important in LWFA driven by longer laser pulses that reach the depletion length at the end of the interaction, where the laser pulse duration is often reduced to a few optical cycles.

The first theoretical treatment of the effect of the CEP on the LWFA process is by Nerush \emph{et al.} \citep{nerush_carrier-envelope_2009}, as a reaction on the plasma bubble oscillations observed in \citep{zhidkov_electron_2008}, initially ascribed to electron-hose instability. Developing a mathematical analysis beyond the ponderomotive model, Nerush \emph{et al.} show that the higher-order terms of the plasma response contain an asymmetry that is CEP dependent, pushing the plasma bubble off-axis. Because of plasma dispersion, the CEP changes with pulse propagation such that the plasma bubble starts to oscillate in the plane of laser polarization. As the electron self-injection region oscillates with the bubble, the electron beam is modulated in a way that depends on the CEP. One can estimate the scale on which the CEP changes by $2\pi$ as \citep{nerush_carrier-envelope_2009,faure_review_2018}:
\begin{equation}
    L_{2\pi}\simeq \frac{n_c}{n_e}\lambda_0
\end{equation}
where $n_e$, $n_c$ and $\lambda_0$ are the electron density, critical electron density and the laser wavelength respectively.

At low intensities, where the plasma bubble does not present any asymmetry, CEP effects can still be significant, as shown by Lifschitz \emph{et al.}\citep{lifschitz_optical_2012} for the case of ionization injection. Here, a varying CEP changes the initial conditions at which the electrons are injected, resulting in different electron energies and emission angles.

Experimentally, a first evidence of a CEP effect in a LWFA experiment is presented in \citep{faure_review_2018}, showing a significant change in the electron spectrum when changing the laser CEP by $\pi/2$. It should be noted that in order to experimentally observe the CEP effect a high level of stability and control of all experimental parameters (pulse energy, pulse duration, beam profile, gas density etc.) is required.

In this paper,  we complement previous work with the motivation of identifying possible observables of CEP effects in realistic experimental contexts where laser wakefield acceleration is driven by near single cycle pulses with moderate intensities (i.e. with normalized vector potential $a_0$ of 3-4). In addition, we emphasize the role of the CEP on the physics of electron injection and acceleration, and focus specifically on self-injection and density gradient (DG) injection. 
The paper is organized as follows: section \ref{sec:asymmetry} presents the CEP-dependent asymmetry of the accelerating structure, and its scaling with intensity. The effect of the CEP on the injection and acceleration of the electron beam is then explained in section \ref{sec:cep_inj_acc}, both for the case of self-injection and density gradient injection. The CEP dependent betatron radiation generated in the self-injection case is analysed in section \ref{sec:betatron}.

\section{Asymmetry of the accelerating structure}\label{sec:asymmetry}
For near single-cycle pulses, the amplitude of the electric field envelope varies considerably over the course of an optical cycle, triggering an asymmetric response of the plasma which depends on the CEP. To quantify this asymmetry, we performed particle-in-cell (PIC) simulations with parameters that correspond to what can realistically be expected in the near future from state-of-the-art near-single-cycle kHz lasers \citep{budriunas_53_2017, ouille_relativistic-intensity_2020}, in combination with high-density, high precision gas jets \citep{salehi_mev_2017,marcinkevicius_femtosecond_2001,tomkus_high-density_2018}.

\subsection{Simulation parameters}
We use the spectral, quasi-cylindrical particle-in-cell code FBPIC \citep{lehe_spectral_2016}, with a modified laser field profile according to \citep{caron_free-space_1999} to accommodate a tightly focused, near-single cycle pulse. With the laser propagating in the z-direction, the radial grid is constructed using $\mathrm{d}z=\lambda_0/50$, $\mathrm{d}r=\lambda_0/20$ (where $\lambda_0=800$ nm is the laser center wavelength), while azimuthally 5 modes are used to correctly capture all departures from cylindrical symmetry. The plasma is made of fully ionized Helium, so that ionization effects are not computed in order to save computation time. In all simulations, the laser pulse duration is 3 fs (FWHM on intensity) and the pulse is focused down to a 5 \si{\micro\m} waist.

\subsection{CEP dependent bubble asymmetry}
\begin{figure*}
    \includegraphics[width=\textwidth]{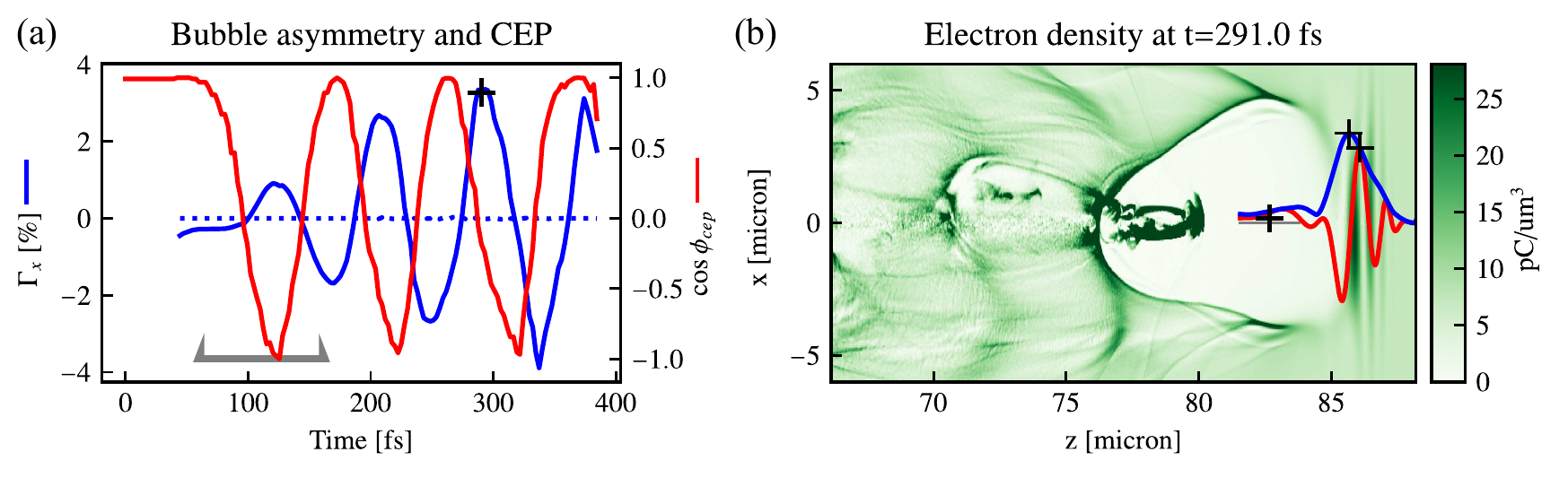}
    \caption{\label{fig:bubble} (a) Asymmetry of the bubble in the plane of polarization (blue, solid) and perpendicular to the polarization (blue, dotted), and the carrier-envelope phase of the driving laser pulse (red), shifting as a result of plasma dispersion. The black dot indicates the position of the snapshot in the right panel. The gray double-headed arrow on the bottom left corresponds to $T_{2\pi}=L_{2\pi}/v_g$. (b) Snapshot of the plasma wave at 291 fs, showing the asymmetric bubble. The black dot indicates the position of the center of mass of the transverse electron density distribution, which is offset with respect to the z-axis (indicated by the gray line) by about 3 \% of $w_0$. The length of the gray line indicates the length over which $\Gamma_x$ is averaged. The electric field and envelope of the laser are drawn in red and blue, respectively, with black dots indicating their maxima (the value of the CEP in this frame is $\pi/2$). }
\end{figure*}

We quantify the bubble asymmetry by calculating the transverse center of mass of the electron density distribution in the plane of polarization ($y=0$), and normalizing it to the bubble radius $R\simeq w_0$:
\begin{equation}\label{eq:asym}
\Gamma_x = \frac{1}{w_0} \frac{\int\limits_X n_e(x)x\,\mathrm{d}x}{\int\limits_X n_e(x)\,\mathrm{d}x},
\end{equation}
where $X$ is the width of the simulation window in $x$. For $a_0=4$ and an electron density $n_e=0.025n_c$ ($n_c$ being the critical electron density 1.74$\times10^{21}$ cm$^{-3}$), figure \ref{fig:bubble} shows how the bubble asymmetry $\Gamma_x$ (blue) oscillates, driven by the CEP of the driving laser pulse (solid red), at an amplitude of 3\% of $w_0$. The panel on the right shows the asymmetric bubble, with the transverse center of mass of the electron density indicated by a black dot. The gray line coincides with the z-axis, and indicates the length over which an average value of $\Gamma_x$ is obtained. This region is chosen there where the asymmetry is apparent and not influenced by the electron beam.
The red and blue lines indicate the laser electric field and its envelope respectively. The CEP is calculated as the distance between the maximum of the electric field and that of the pulse envelope (each indicated by a black spot), divided by the center wavelength of the pulse. The tail of the pulse undergoes a strong redshift, causing the center wavelength to be redshifted by a factor of 2 over the course of the simulation (see Supplementary Information), an effect that was observed in \citep{beaurepaire_electron_2014}. 
Through this redshift, the effective normalized laser field strength $a_0\propto\sqrt{I\lambda^2}$ increases considerably, peaking at 6.8 around $t=240$ fs. As a result, the asymmetry of the plasma response increases over the course of the simulation, until depletion of the laser pulse becomes significant for times $t>300$ fs.
Interestingly, the increase in $a_0$ also increases the transverse momentum of the electrons ejected by the laser pulse, which causes the bubble top and bottom parts to flatten (see also supplementary movie). These effects cause the bubble oscillation frequency to slightly decrease, as can be seen in figure \ref{fig:bubble} in the decreasing period of $\Gamma_x$ (blue curve).

In the plane perpendicular to the laser polarization ($x=0$) the bubble asymmetry is absent, as shown by the dotted blue line in figure \ref{fig:bubble}. One may wonder what happens in the case of circular polarization - as shown in the Supplementary Information, the bubble and electron beam then describe a (CEP dependent) rotary motion behind the laser pulse.

\subsection{Intensity scaling of the bubble asymmetry}
\begin{figure}
    \includegraphics[width=\columnwidth]{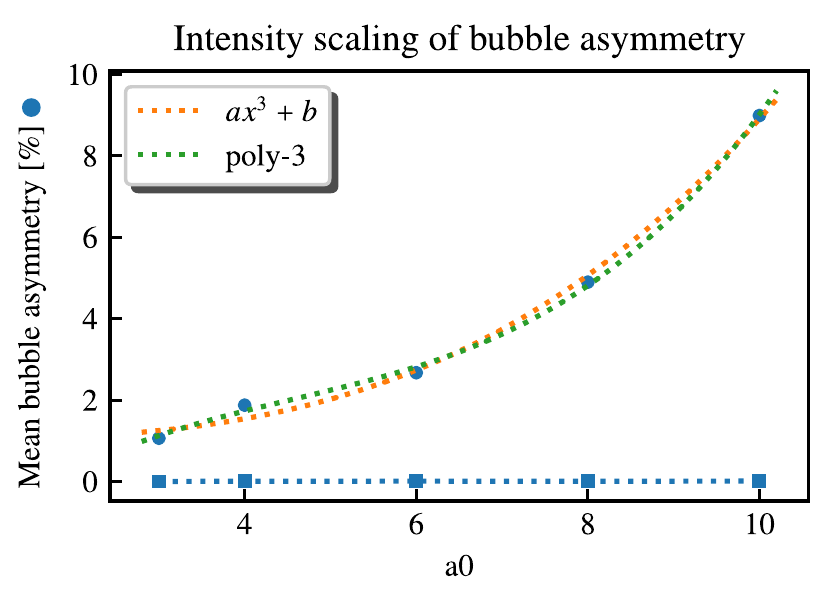}
    \caption{\label{fig:intensityscaling}Mean bubble asymmetry increases as a function of vacuum laser field strength. According to \cite{nerush_carrier-envelope_2009} the asymmetry scales with $a_0^3$. We have thus fitted the bubble asymmetry with a cubic function (orange) and a 3rd order polynomial (green). The bubble asymmetry in the perpendicular plane remains zero as expected.}
\end{figure}

The strength of the asymmetry is expected to depend on the laser intensity. We repeated the simulation for (vacuum) laser field strengths ranging from $a_0=3$ to $a_0=10$, keeping all other parameters constant, and calculated a mean bubble asymmetry by averaging $|\Gamma|$ from the moment the bubble is formed up to the end of the plasma. Figure \ref{fig:intensityscaling} shows this mean bubble asymmetry as a function of the laser intensity. It is interesting to compare this to the predictions from Nerush \emph{et al.} \citep{nerush_carrier-envelope_2009}. According to the authors, the asymmetry becomes considerable only for $a_0>k_0w_0$ (where $k_0=\omega_0/c$ is the laser wave vector), i.e. for $a_0>12$ for our set of parameters ($\lambda_0=800$ nm and $w_0=5$ \si{\micro\m}), but we clearly observe a strong asymmetry at lower laser intensities. This can be explained by the strong redshift of the laser pulse: as the center wavelength increases by a factor 2, the theoretically predicted limit drops to $a_0>6$, which is indeed reached by the laser field strength when normalized to the redshifted laser wavelength.

Secondly, the bubble asymmetry in \citep{nerush_carrier-envelope_2009} is described to scale with $a_0^3$. Fitting a cubic function ($ax^3+b$) yields the orange curve in the right panel of figure \ref{fig:intensityscaling}, while including the full third order polynomial (green) improves the fit only marginally. Despite the fact that the definition of the bubble asymmetry in \citep{nerush_carrier-envelope_2009} slightly differs from ours \footnote{In \citep{nerush_carrier-envelope_2009} bubble asymmetry $\Lambda (l)$ is defined as the sum of transverse momenta of two electrons born with initial position $+l$ and $-l$ in the polarization plane.}, the agreement with the $a_0
^3$ scaling is remarkable. 

In summary, our results are consistent with the predictions from \citep{nerush_carrier-envelope_2009} concerning the cubic scaling, but the bubble asymmetry appears to be considerable even at intensities lower than one may expect from the theoretical prediction, due to the strong redshift.

\section{CEP-effects on electron injection and acceleration}\label{sec:cep_inj_acc}
In a pulse so short that the ponderomotive approximation is no longer valid, the changing CEP influences not only the accelerating structure, but also the electron beam injection and acceleration. We will now show that in the case of self-injection, it is the asymmetry at the moment of injection that governs the CEP-dependent dynamics of the electron beam (rather than the oscillating bubble during acceleration).

\subsection{Self-injection}\label{sec:selfinjection}
Possibly because self-injection is easy to implement experimentally, it has been currently the most widespread injection mechanism in LWFA since its demonstration over two decades ago \citep{modena_electron_1995,umstadter_nonlinear_1996}. It is based on a sudden increase of the bubble size as the laser pulse reaches highly relativistic intensities. As described above, electrons at the front of the bubble collectively gain a transverse momentum which is CEP-dependent. As these electrons travel towards the back of the bubble, the transverse momentum changes sign and at the back of the bubble, the electrons are injected and start describing a collective betatron oscillation \citep{rousse_production_2004}. When the injection length is small compared to $L_{2\pi}$, the CEP at the moment of injection can thus strongly influence the electron beam produced by the accelerator. In the opposite case, the effects may be averaged out over an oscillation of the CEP.

\begin{figure}
    \includegraphics[width=\columnwidth]{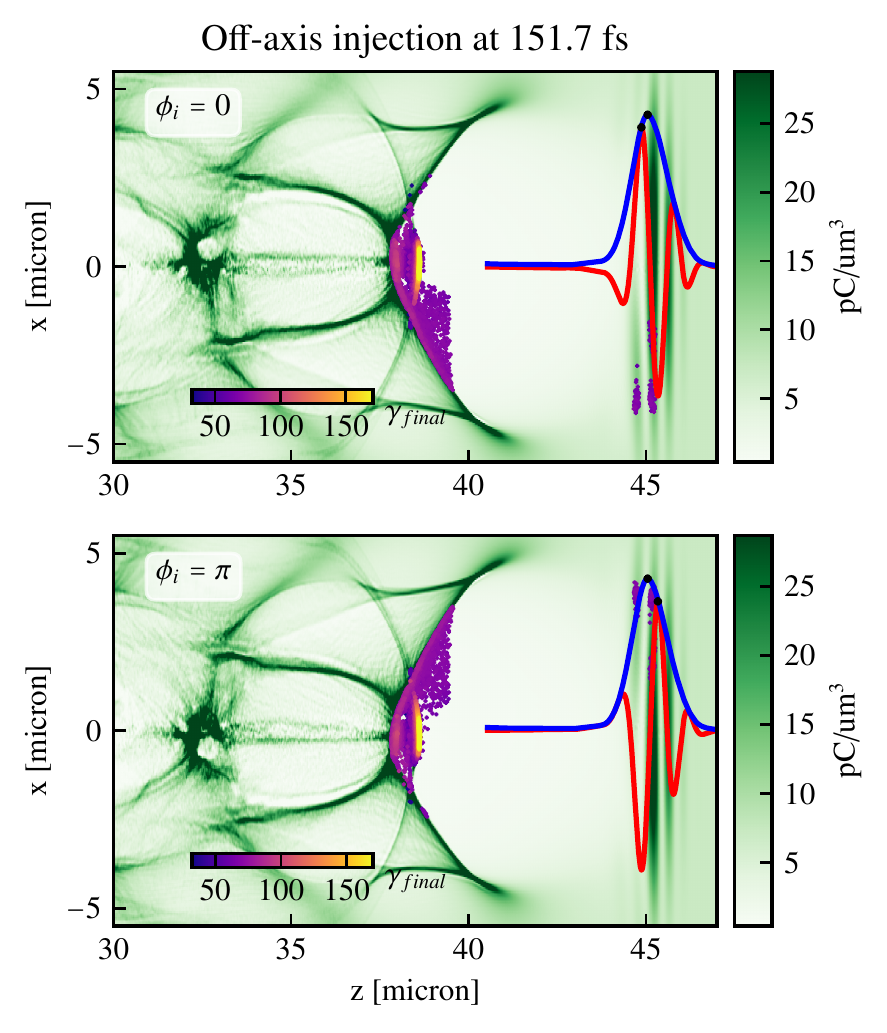}
    \caption{\label{fig:offaxisinjection}Simulation snapshots at $t=151.7$ fs, for initial CEP values $\phi_i=0$ and $\phi_i=\pi$, showing the electron density in shades of green and the injected electrons that constitute the electron beam. The color of the injected electrons corresponds to their final energy $\gamma_{final}$. These two frames clearly show the off-axis injection of the electron beam. High energy electrons are injected on-axis, whereas off-axis injected electrons form a lower-energy tail performing strong betatron oscillation. See Supplementary Material for the full movie.}
\end{figure}

To illustrate this effect, we study self-injection in the simulation of figure \ref{fig:bubble} (i.e. $a_0=4$, $n_e=0.025n_c$) for a laser pulse with an initial CEP of $\phi_i=0$ and $\phi_i=\pi$. Figure \ref{fig:offaxisinjection} shows the moment of self-injection for these two cases. In these frames, the electron density is shown in shades of green, and the injected electrons that will constitute the accelerated electron beam are shown in a purple-yellow color scale corresponding to their final energy $\gamma_{final}$. The asymmetric nature of the injection is evident and the color scale shows how the most energetic electrons are injected nearly on-axis, whereas the lower energy electrons are injected off-axis. As we will see below, this lower-energy tail subsequently performs large amplitude collective betatron oscillations during acceleration. Note that in the common case of longer pulses, the ponderomotive approximation still holds and the laser pulse drives a symmetric plasma bubble. Electrons injected in this symmetric structure also perform betatron oscillations, but individually, i.e. the ensemble average of the beam is symmetric in $x$ and in $y$. What we see here is a different phenomenon, i.e. a collective betatron oscillation of the beam, which is CEP dependent. As noted in \citep{nerush_carrier-envelope_2009} these CEP-dependent oscillations are also easily distinguished from beam-hose instabilities \citep{whittum_electron-hose_1991} as such instabilities are not confined to the plane of laser polarization, in addition to the fact that figure \ref{fig:bubble} clearly shows that the oscillation is locked to the CEP of the driving laser.
To visualize this process in greater detail the reader is encouraged to explore the movies in the Supplementary Material.

\begin{figure}
    \includegraphics[width=\columnwidth]{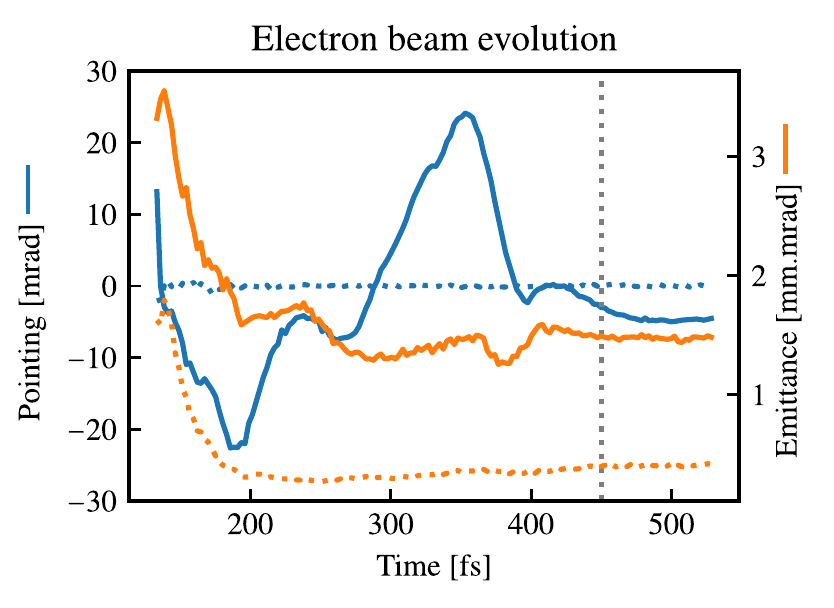}
    \caption{\label{fig:a04_cep0_pointingemittance}Beam pointing (blue) and transverse normalized emittance (orange) in the plane of laser polarization (solid) and the perpendicular plane (dotted). The effects of the oscillating CEP and bubble asymmetry are most evident on the pointing of the electron beam.}
\end{figure}

We will now track the development of the electron beam throughout the acceleration. In order to get meaningful results a proper definition of ``the electron beam'' is crucial. We define the electron beam as: the electrons which in the final iteration (when the electron beam has left the plasma) have forward momentum $u_z>u_{z,min}$ and lie within a 50 mrad angle of the most energetic part of the beam. Based on the electron beam's phase space distribution, $u_{z,min}$ is the threshold which separates the electrons in the first accelerating bucket from the background electrons, in this case corresponding to $u_z>u_{z,min}=26m_e c$. 
Following these electrons, we can track the evolution of the beam pointing and transverse normalized emittance, plotted in blue and orange in figure \ref{fig:a04_cep0_pointingemittance} for an initial CEP of $\phi_i=0$ (the evolution of the beam charge and energy can be found in the Supplementary Information). Of these beam parameters the pointing in particular shows a clear oscillation around the symmetry axis in the plane of the laser polarization, which is the collective betatron oscillation described above. Incidentally, a similar effect was observed in \citep{glinec_direct_2008}, where collective betatron oscillations following asymmetric injection was caused by an asymmetric transverse laser profile, or in \citep{popp_all-optical_2010} where a laser pulse-front tilt was used to influence beam pointing. The reader is reminded of the characteristic frequency of betatron oscillation for an electron in a focusing wakefield, which is given by:
\begin{equation}
    \omega_{\beta} = \frac{\omega_p}{\sqrt{2\gamma}},
\end{equation}
where $\omega_p$ is the plasma frequency. 
In the perpendicular plane (dotted line) no significant oscillation is observed, as expected. The evolution of the total beam charge shows that self-injection takes place from 110 to 200 fs into the simulation, which corresponds to a length of 25 \si{\micro\m}, or $0.8L_{2\pi}$. Different parts of the electron beam are thus injected at a different phase of the bubble oscillation, resulting in a superposition of multiple oscillations in the pointing of the electron beam. In addition, as the electrons gain energy from 6 to 44 MeV, their betatron oscillation period increases roughly linearly (in time) from 90 fs to 225 fs. These values are consistent with the oscillations observed in figure \ref{fig:a04_cep0_pointingemittance}.
The emittance of the electron beam is significantly larger in the plane of polarization than in the perpendicular plane by a factor 3.6. Again, this is an effect that cannot be explained within the ponderomotive approximation, but is a direct consequence of the injection asymmetry. Note that although asymmetric emittances have been reported before, this was mostly due to the electron beam interacting with the rear of the laser pulse \citep{phuoc_imaging_2006}. In our simulation, electrons are injected through pure self-injection, and the observed effect can be attributed solely to the asymmetry of the plasma response in the plane of polarization.

\begin{figure*}
    \includegraphics[width=\textwidth]{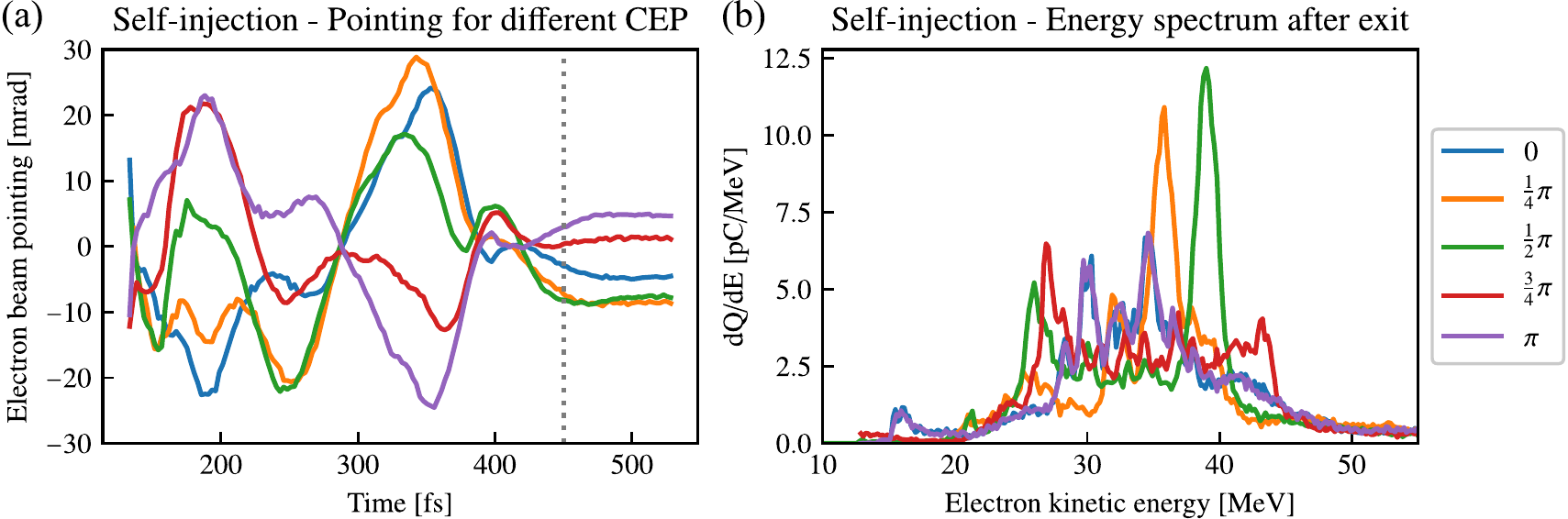}
    \caption{\label{fig:cepscan_a04}(a) Pointing of the electron beam as a function of time, for initial CEP values from 0 to $\pi$. Note the exact symmetry between the trajectories 0 and $\pi$. The gray dotted line indicates the plasma-vacuum interface. (b) The final electron beam energy spectra for the different CEP values. The spectra for 0 and $\pi$ overlap.}
\end{figure*}

We would now like to see what happens when the initial CEP of the laser pulse is varied, and how it affects the electron beam pointing, as could be observed in an experiment. We repeated the same simulation with five different initial CEP values ranging from 0 to $\pi$. The evolution of the beam pointing for these five cases (figure \ref{fig:cepscan_a04}, left) clearly shows the importance of controlling the CEP to obtain a stable accelerator. As one would expect, the cases of $0$ and $\pi$ are symmetric with respect to the z-axis. The pointing deviation at the exit of the plasma (grey dotted line) is on the order of 10 mrad, and depends strongly on the length of the plasma. Indeed, in simulations with a shorter gas jet length of 50 \si{\micro\m} (instead of 80, corresponding to the dephasing length) the electron beam leaves the plasma after 350 fs with a beam pointing angle of up to 20 mrad. Thus, in an experiment that aims at measuring CEP effects on the electron beam, it is crucial to keep a highly stable gas density profile.

The different phase-space trajectories followed by the electron beams for different CEP values have a significant influence on their energy spectrum, as is clear from figure \ref{fig:cepscan_a04} (right panel). The spectra for initial CEP values 0 and $\pi$ overlap. Again, the importance of controlling the CEP is evident: an initial CEP of $\pi/2$ gives a quasi-mono-energetic peak of 5 \% at 39 MeV, as opposed to a broad 25 \% energy spectrum at a CEP of 0 or $\pi$.

\begin{figure}
    \includegraphics[width=\columnwidth]{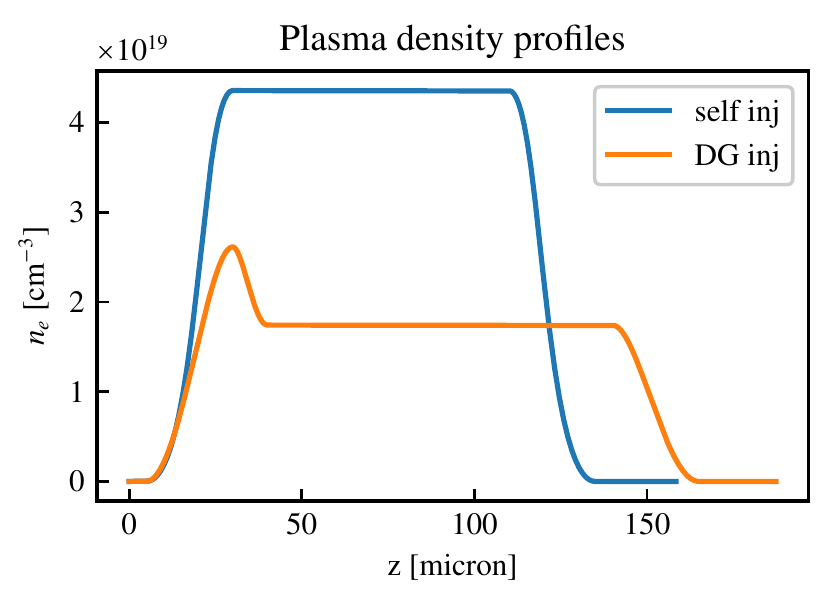}
    \caption{The electron density profiles corresponding to the simulations for the case of self injection (blue) and density gradient injection (orange).}\label{fig:densityprofiles}
\end{figure}

\subsection{Density gradient injection}\label{sec:dginjection}
A second injection mechanism well-adapted to LWFA with near-single-cycle pulses is injection in a density gradient (DG) \citep{bulanov_particle_1998, geddes_plasma-density-gradient_2008,schmid_density-transition_2010,beaurepaire_effect_2015}. It consists in tailoring the gas jet such that the density profile features a sharp downward gradient, causing a sudden increase of the plasma wavelength. The back of the plasma bubble thus undergoes an effective slow-down which triggers trapping of electrons.
Where self-injection is triggered by the highly nonlinear response of the plasma to an ultrarelativistic laser pulse, injection in a density gradient is controlled by the relatively stable density profile. It thus allows for more controlled injection, at lower intensities. If the density gradient is over a length scale significantly smaller than $L_{2\pi}$, injection takes place at a specific CEP value and an effect on the electron beam can be expected. However, the injection geometry is to a first approximation determined by the gas density gradient, so one can expect the effect to be significantly weaker than in the case of self-injection. A similar effect, albeit with different underlying physics, is described in \citep{corde_observation_2013} comparing transverse and longitudinal injection in laser-plasma accelerators. In short, self-injection is typically dominated by transverse injection, i.e. electrons have are injected with an initial transverse momentum. In the case of longitudinal injection electrons remain essentially on-axis and are injected with virtually no transverse momentum, yielding an accelerator that is more robust to changes in laser profile. Similarly, in our example of DG-injection, electrons are injected with low initial transverse momentum and the resulting electron beam is more robust to changes in the CEP.

Because DG-injection is well adapted for applications with limited laser pulse energy, and in order to avoid self-injection, we performed the DG simulations at a (vacuum) laser field strength of $a_0=3$. The peak electron density at the start of the gas jet is $0.015n_c$ which drops to $0.01n_c$ over a length of 10 \si{\micro\m} (see figure \ref{fig:densityprofiles}).

\begin{figure}
    \includegraphics[width=\columnwidth]{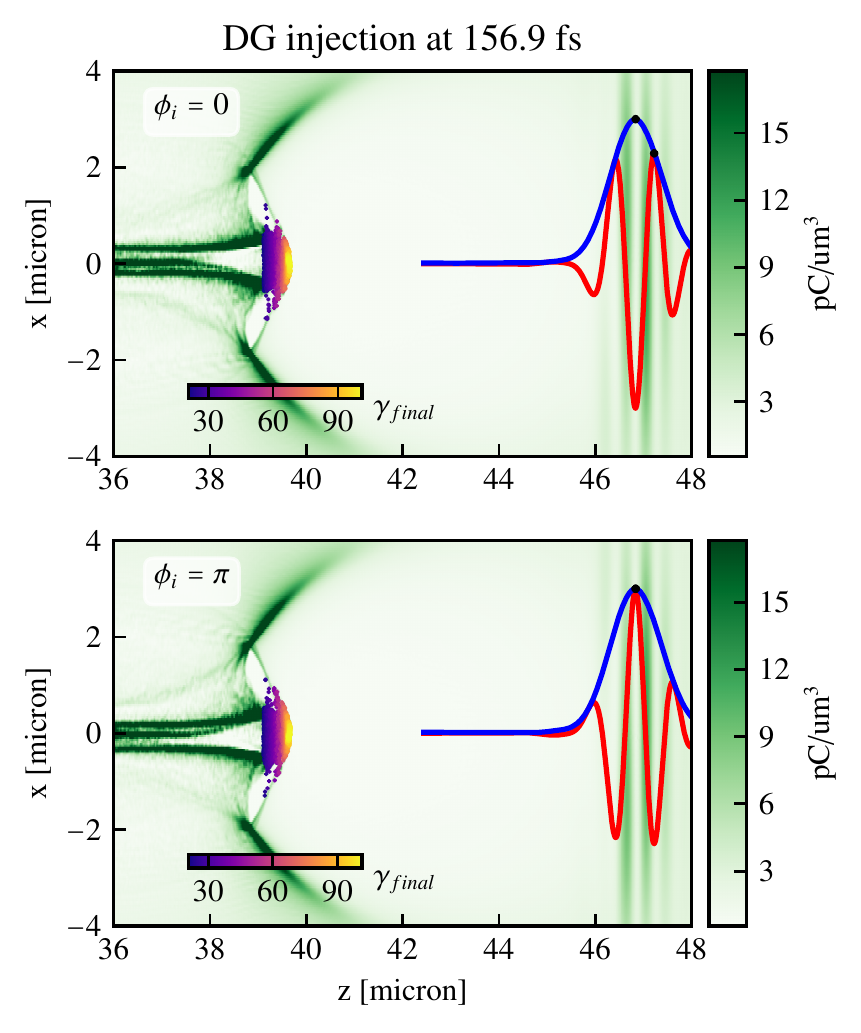}
    \caption{\label{fig:dginjection}Simulation snapshots at $t=156.9$ fs, for initial CEP values 0 and $\pi$, showing the electron density in shades of green and the injected electrons that constitute the electron beam. The color of the injected electrons corresponds to their final energy $\gamma_{final}$. These two frames clearly show how the injection is axisymmetric to a first approximation.}
\end{figure}

As in figure \ref{fig:offaxisinjection} for the case of self-injection, figure \ref{fig:dginjection} shows the two frames corresponding to the moment of electron beam injection (t=156.9 fs) in the density gradient, for initial CEP values of 0 and $\pi$. At first sight, the injection is indeed highly axisymmetric and, to a first approximation, independent of the CEP, in line with the above-mentioned reasoning that the injection is determined by the geometry of the gas density profile (i.e. longitudinal).

\begin{figure*}
    \includegraphics[width=\textwidth]{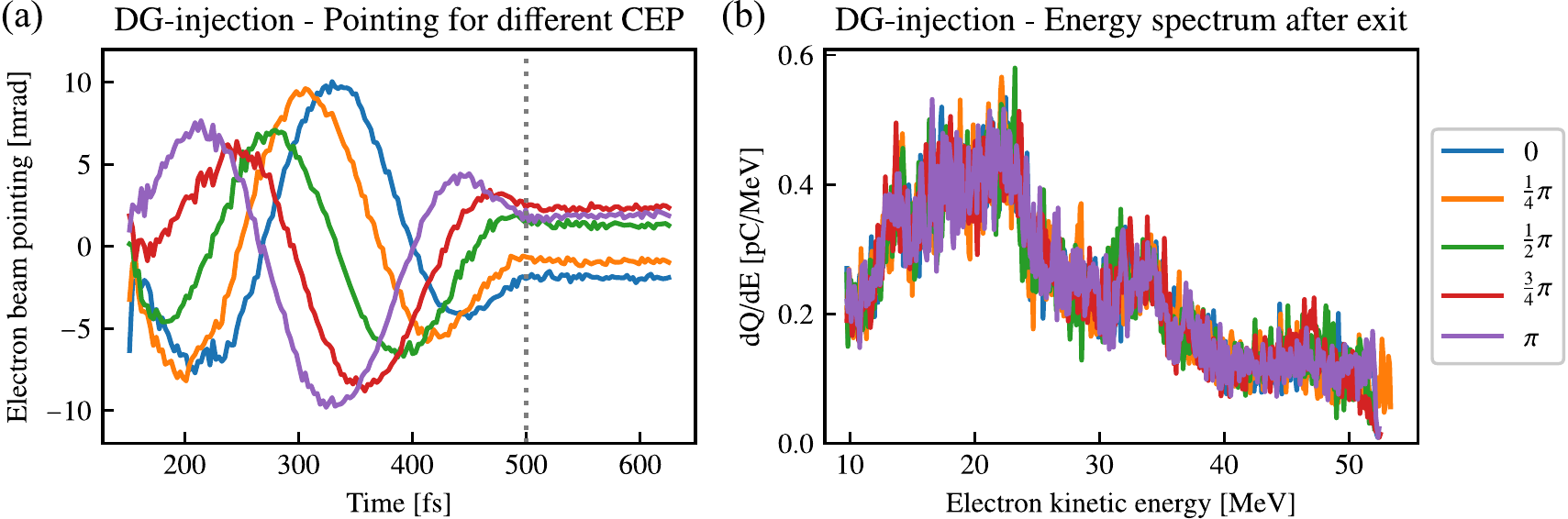}
    \caption{\label{fig:cepscan_dg}Injection in a density gradient. (a) Pointing of the electron beam as a function of time, for initial CEP values from 0 to $\pi$. Again, the trajectories for 0 and $\pi$ are symmetric with respect to the z-axis. The gray dotted line indicates the plasma-vacuum interface. (b) The final electron beam energy spectra for the different CEP values.}
\end{figure*}

An analysis of the beam pointing plotted in figure \ref{fig:cepscan_dg} shows however that there is still a betatron oscillation in the electron beam, meaning the injection is not perfectly axisymmetric. As in the case of self-injection, the electrons that end up in the beam have collectively received an initial kick from the laser field (positive $x$ for $\phi_i=0$, negative $x$ for $\phi_i=\pi$) which causes a small amplitude betatron oscillation (compare also the movies in the Supplementary Information). The oscillation is cleaner than in the case of self-injection as the electron injection is highly localized (see evolution of the beam charge in the Supplementary Information). In addition, the oscillation amplitude is a factor 2 smaller than for the self-injection case at $a_0=3$ (not shown). The average bubble asymmetry is 1.1 \%, similar to the self-injection case at $a_0=3$. In general, bubble asymmetry was found to be independent of electron density.

Over the course of the simulation, as the electrons gain energy from about 12 to 22 MeV, the calculated betatron oscillation period increases from 190 to 250 fs, which is consistent with the oscillation period in figure \ref{fig:cepscan_dg}. 

Importantly, we see that contrary to the self-injection case, the electron beam energy spectrum (right) is independent of the CEP. This is due to the fact that injection is governed by the plasma density and strictly localized to the same location, independent of the CEP. When comparing these results to those of self-injection, it is important to keep in mind that for the simulations for density gradient injection the laser intensity and electron density were lowered to avoid self-injection and that therefore these results cannot be compared one-to-one. The results however illustrate the fundamental difference between the two injection mechanisms.

\section{Betatron radiation}\label{sec:betatron}
\begin{figure*}
    \includegraphics[width=\textwidth]{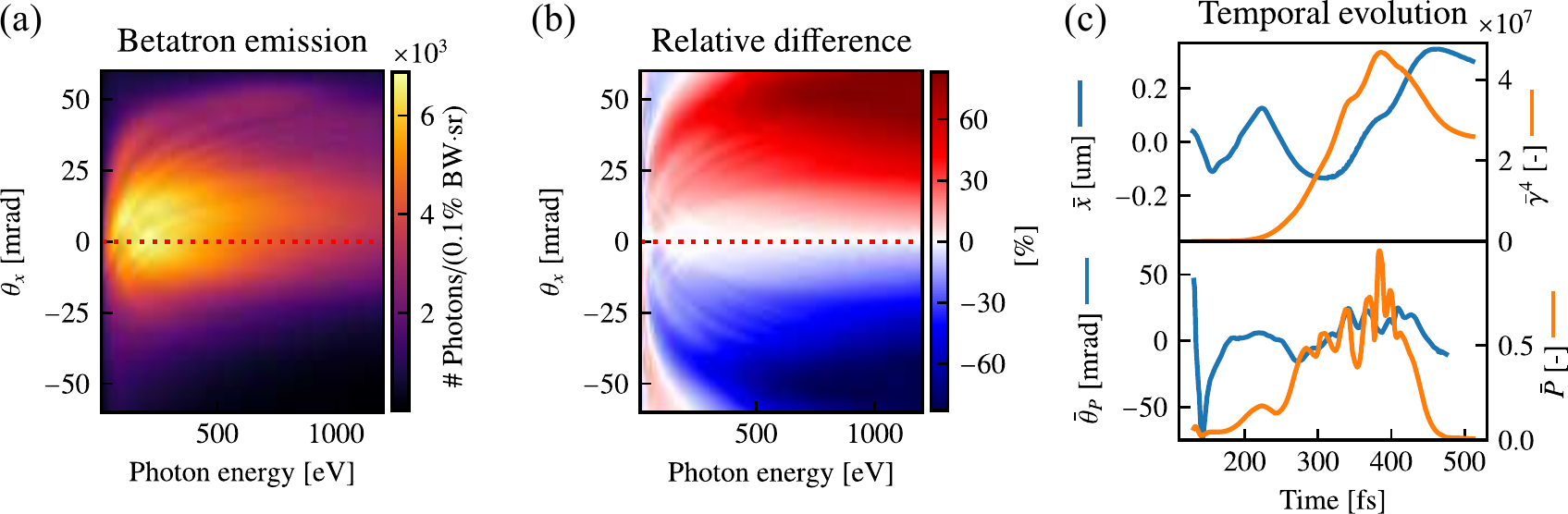}
    \caption{\label{fig:betatronasymm} Asymmetry of the emitted betatron radiation for the self-injection case, $a_0=4$. (a) Angularly resolved betatron spectrum for an initial CEP of $\phi_i=0$. (b) The difference between the spectrum of (a) and the same spectrum for $\phi_i=\pi$, normalized to the sum of the two spectra. The dotted red lines at $\theta_x=0$ are a guide to the eye. (c) Upper panel: In blue, average transverse coordinate of all electrons that contribute to betatron radiation ($\bar{x}$). In orange, the average value of $\gamma^4$ for these electrons. Lower panel: In blue, the average angle $\bar{\theta}_P$ of the trajectories fo the emitting electrons, weighted by the relative power of the emitters at each timestep. In orange, the power $P$ averaged over all contributing electrons, normalized to its maximum value.}
\end{figure*}

The clear dependence of the betatron oscillations on the initial CEP that is observed in the case of self-injection (figure \ref{fig:cepscan_a04}) suggests that the emitted betatron radiation should also be CEP-dependent. By evaluating the Liénard-Wiechert potentials \citep{jackson_classical_1999,andriyash_synchrad}, the betatron radiation was calculated for initial CEP values of 0 and $\pi$, for self-injection and density gradient injection (see Supplementary Information for the density gradient case). As shown in figure \ref{fig:betatronasymm} (a) and (b), the generated betatron radiation shows a strong asymmetry in the plane of polarization. Such a clear CEP dependence of the emitted betatron radiation is directly observable in experiments. It can be explained by looking at the electron trajectories. We recall that the radiated power $P_\gamma$ is proportional to:
\begin{equation}
P_\gamma \propto \frac{\gamma^4}{\rho^2}, 
\end{equation}
where $\rho$ is the radius of curvature of the electron trajectory \citep{jackson_classical_1999,corde_femtosecond_2013}. Betatron radiation is thus emitted at points where the electron trajectory has a small radius of curvature and thus a fast change in beam pointing (i.e. a large transverse acceleration) and is entirely dominated by regions of high $\gamma$. As the upper panel of figure \ref{fig:betatronasymm} c shows, the evolution of $\gamma$ causes betatron emission to be dominated by the contribution of the last half-cycle of the betatron oscillation. As the electron beam is pointing in the positive $x$-direction during this last half-cycle, the observed betatron radiation is shifted towards positive $\theta_x$. A more precise evaluation of the betatron emission direction can be given by looking at the angle $\theta_x$ of each electron trajectory, and computing the average angle ($\bar{\theta}_P$) of betatron emission by weighting the trajectories by their radiated power at each timestep (lower panel of figure \ref{fig:betatronasymm}). It shows that at timepoints where the radiated power is high (ensemble averaged power $\bar{P}$, lower panel in orange), the emitting electrons are pointing in the range $-10$ to 25 mrad, which is coherent with panels (a) and (b).

Note that CEP-effects on the emitted betatron radiation are observable only because the acceleration is short, such that the electron beam only radiates significantly during half a betatron oscillation. If the beam described a full or even multiple oscillations at a near-constant $\gamma$, CEP-effects would be averaged out. With near single-cycle pulses short acceleration lengths are inevitable, as the acceleration is limited by depletion of the laser. Indeed the depletion length $L_{depl}=\frac{\omega_0^2}{\omega_p^2}c\tau=36$ \si{\micro\m} is about half the betatron wavelength $\lambda_{\beta}=\frac{2\pi c}{\omega_p}\sqrt{2\gamma}=64$ \si{\micro\m} for our parameters (see also the temporal evolution of the laser field strength in the Supplementary Information).

\section{Conclusion}
Through PIC simulations we explored the effect of the CEP on injection and acceleration in a laser-plasma accelerator, in particular the cases of self-injection and density gradient injection with near-single cycle laser pulses at moderately relativistic laser intensities. For such short pulses, the ponderomotive approximation is no longer valid and the plasma response becomes asymmetric in the plane of laser polarization. In contrast to what one may expect from theoretical predictions \citep{nerush_carrier-envelope_2009}, bubble asymmetry is found to be significant even at these moderate intensities due to significant redshifting of the laser pulse. The observed oscillation of the bubble is locked to the CEP which changes with propagation through plasma dispersion. As the laser redshifts during propagation, $a_0$ increases up to 6.8 (compared to 4.0 in vacuum), causing an increasingly asymmetric plasma response. For the case of self-injection, the transverse momentum gained by the electrons causes them to be injected off-axis and subsequently launches large amplitude collective betatron oscillations. While such large oscillations may be beneficial to the generation of betatron X-ray radiation, they also cause a degradation of the transverse emittance as well as oscillations of the beam pointing along propagation. This collective betatron oscillation depends on the betatron frequency and on the extent to which the injection is localized in time (if the injection time is more than a fraction of the CEP period, a superposition of several oscillations is observed). The angle at which the electron beam leaves the accelerator is thus influenced by the initial CEP of the laser. The energy spectrum of the electron beam is also shown to be strongly CEP-dependent. The emitted betatron radiation offers another CEP-dependent observable: due to the short acceleration length and the strong dependence of the emission on $\gamma$, betatron emission is limited to the last half oscillation and its pointing thus strongly depends on the CEP. These effects are reduced in the case of injection in a density gradient, where injection is axisymmetric to a first approximation and the electron energy spectrum is independent of the CEP.

These findings show to which extent CEP control, especially in the case of self-injection, is crucial in establishing a laser wakefield accelerator capable of producing relativistic electron bunches using near single-cycle pulses, with a stable and reproducible beam pointing and energy spectrum. In future experiments, controlling the CEP may become a way of controlling the energy spectrum and pointing of the electron beam.
Alternatively, using a density gradient injection scheme loosens the requirements on the CEP stability of the laser, which is relevant for systems where CEP-stabilization is challenging or absent.
These findings are an important step in the maturation of these unique sources.

\section{Acknowledgements}
The authors acknowledge fruitful discussions with Neil Zaïm and Remi Lehe on the implementation of near single-cycle pulses in FBPIC.


\begin{thebibliography}{42}
\providecommand{\natexlab}[1]{#1}
\providecommand{\url}[1]{\texttt{#1}}
\expandafter\ifx\csname urlstyle\endcsname\relax
  \providecommand{\doi}[1]{doi: #1}\else
  \providecommand{\doi}{doi: \begingroup \urlstyle{rm}\Url}\fi

\bibitem[Tajima and Dawson(1979)]{tajima_laser_1979}
T.~Tajima and J.~M. Dawson.
\newblock Laser {Electron} {Accelerator}.
\newblock \emph{Phys. Rev. Lett.}, 43\penalty0 (4):\penalty0 267--270, July
  1979.
\newblock \doi{10.1103/PhysRevLett.43.267}.
\newblock URL \url{https://link.aps.org/doi/10.1103/PhysRevLett.43.267}.
\newblock Publisher: American Physical Society.

\bibitem[Esarey et~al.(2009)Esarey, Schroeder, and
  Leemans]{esarey_physics_2009}
E.~Esarey, C.~B. Schroeder, and W.~P. Leemans.
\newblock Physics of laser-driven plasma-based electron accelerators.
\newblock \emph{Rev. Mod. Phys.}, 81\penalty0 (3):\penalty0 1229--1285, August
  2009.
\newblock ISSN 0034-6861, 1539-0756.
\newblock \doi{10.1103/RevModPhys.81.1229}.
\newblock URL \url{https://link.aps.org/doi/10.1103/RevModPhys.81.1229}.

\bibitem[Faure et~al.(2004)Faure, Glinec, Pukhov, Kiselev, Gordienko, Lefebvre,
  Rousseau, Burgy, and Malka]{faure_laserplasma_2004}
J.~Faure, Y.~Glinec, A.~Pukhov, S.~Kiselev, S.~Gordienko, E.~Lefebvre, J.-P.
  Rousseau, F.~Burgy, and V.~Malka.
\newblock A laser–plasma accelerator producing monoenergetic electron beams.
\newblock \emph{Nature}, 431\penalty0 (7008):\penalty0 541--544, September
  2004.
\newblock ISSN 0028-0836, 1476-4687.
\newblock \doi{10.1038/nature02963}.
\newblock URL \url{http://www.nature.com/articles/nature02963}.

\bibitem[Mangles et~al.(2004)Mangles, Murphy, Najmudin, Thomas, Collier,
  Dangor, Divall, Foster, Gallacher, Hooker, Jaroszynski, Langley, Mori,
  Norreys, Tsung, Viskup, Walton, and Krushelnick]{mangles_monoenergetic_2004}
S.~P.~D. Mangles, C.~D. Murphy, Z.~Najmudin, A.~G.~R. Thomas, J.~L. Collier,
  A.~E. Dangor, E.~J. Divall, P.~S. Foster, J.~G. Gallacher, C.~J. Hooker,
  D.~A. Jaroszynski, A.~J. Langley, W.~B. Mori, P.~A. Norreys, F.~S. Tsung,
  R.~Viskup, B.~R. Walton, and K.~Krushelnick.
\newblock Monoenergetic beams of relativistic electrons from intense
  laser–plasma interactions.
\newblock \emph{Nature}, 431\penalty0 (7008):\penalty0 535--538, September
  2004.
\newblock ISSN 0028-0836, 1476-4687.
\newblock \doi{10.1038/nature02939}.
\newblock URL \url{http://www.nature.com/articles/nature02939}.

\bibitem[Geddes et~al.(2004)Geddes, Toth, van Tilborg, Esarey, Schroeder,
  Bruhwiler, Nieter, Cary, and Leemans]{geddes_high-quality_2004}
C.~G.~R. Geddes, Cs. Toth, J.~van Tilborg, E.~Esarey, C.~B. Schroeder,
  D.~Bruhwiler, C.~Nieter, J.~Cary, and W.~P. Leemans.
\newblock High-quality electron beams from a laser wakefield accelerator using
  plasma-channel guiding.
\newblock \emph{Nature}, 431\penalty0 (7008):\penalty0 538--541, September
  2004.
\newblock ISSN 0028-0836, 1476-4687.
\newblock \doi{10.1038/nature02900}.
\newblock URL \url{http://www.nature.com/articles/nature02900}.

\bibitem[Lundh et~al.(2011)Lundh, Lim, Rechatin, Ammoura, Ben-Ismaïl, Davoine,
  Gallot, Goddet, Lefebvre, Malka, and Faure]{lundh_few_2011}
O.~Lundh, J.~Lim, C.~Rechatin, L.~Ammoura, A.~Ben-Ismaïl, X.~Davoine,
  G.~Gallot, J-P. Goddet, E.~Lefebvre, V.~Malka, and J.~Faure.
\newblock Few femtosecond, few kiloampere electron bunch produced by a
  laser–plasma accelerator.
\newblock \emph{Nature Phys}, 7\penalty0 (3):\penalty0 219--222, March 2011.
\newblock ISSN 1745-2473, 1745-2481.
\newblock \doi{10.1038/nphys1872}.
\newblock URL \url{http://www.nature.com/articles/nphys1872}.

\bibitem[Ta~Phuoc et~al.(2008)Ta~Phuoc, Corde, Fitour, Shah, Albert, Rousseau,
  Burgy, Rousse, Seredov, and Pukhov]{ta_phuoc_analysis_2008}
Kim Ta~Phuoc, Sebastien Corde, Romuald Fitour, Rahul Shah, Felicie Albert,
  Jean-Philippe Rousseau, Fréderic Burgy, Antoine Rousse, Vasily Seredov, and
  Alexander Pukhov.
\newblock Analysis of wakefield electron orbits in plasma wiggler.
\newblock \emph{Physics of Plasmas}, 15\penalty0 (7):\penalty0 073106, July
  2008.
\newblock ISSN 1070-664X.
\newblock \doi{10.1063/1.2952831}.
\newblock URL \url{https://aip.scitation.org/doi/10.1063/1.2952831}.
\newblock Publisher: American Institute of Physics.

\bibitem[He et~al.(2013)He, Hou, Nees, Easter, Faure, Krushelnick, and
  Thomas]{he_high_2013}
Z.-H. He, B.~Hou, J.~A. Nees, J.~H. Easter, J.~Faure, K.~Krushelnick, and
  A.~G.~R. Thomas.
\newblock High repetition-rate wakefield electron source generated by
  few-millijoule, 30 fs laser pulses on a density downramp.
\newblock \emph{New J. Phys.}, 15\penalty0 (5):\penalty0 053016, May 2013.
\newblock ISSN 1367-2630.
\newblock \doi{10.1088/1367-2630/15/5/053016}.
\newblock URL \url{https://doi.org/10.1088\%2F1367-2630\%2F15\%2F5\%2F053016}.

\bibitem[Mahieu et~al.(2018)Mahieu, Jourdain, Phuoc, Dorchies, Goddet,
  Lifschitz, Renaudin, and Lecherbourg]{mahieu_probing_2018}
B.~Mahieu, N.~Jourdain, K.~Ta Phuoc, F.~Dorchies, J.-P. Goddet, A.~Lifschitz,
  P.~Renaudin, and L.~Lecherbourg.
\newblock Probing warm dense matter using femtosecond {X}-ray absorption
  spectroscopy with a laser-produced betatron source.
\newblock \emph{Nat Commun}, 9\penalty0 (1):\penalty0 1--6, August 2018.
\newblock ISSN 2041-1723.
\newblock \doi{10.1038/s41467-018-05791-4}.
\newblock URL \url{https://www.nature.com/articles/s41467-018-05791-4}.
\newblock Number: 1 Publisher: Nature Publishing Group.

\bibitem[Malka et~al.(2008)Malka, Faure, Gauduel, Lefebvre, Rousse, and
  Phuoc]{malka_principles_2008}
Victor Malka, Jérôme Faure, Yann~A. Gauduel, Erik Lefebvre, Antoine Rousse,
  and Kim~Ta Phuoc.
\newblock Principles and applications of compact laser–plasma accelerators.
\newblock \emph{Nature Phys}, 4\penalty0 (6):\penalty0 447--453, June 2008.
\newblock ISSN 1745-2481.
\newblock \doi{10.1038/nphys966}.
\newblock URL \url{https://www.nature.com/articles/nphys966}.
\newblock Number: 6 Publisher: Nature Publishing Group.

\bibitem[Behm et~al.(2020)Behm, Hussein, Zhao, Baggott, Cole, Hill,
  Krushelnick, Maksimchuk, Nees, Rose, Thomas, Watt, Wood, Yanovsky, and
  Mangles]{behm_demonstration_2020}
K.~Behm, A.E. Hussein, T.Z. Zhao, R.A. Baggott, J.M. Cole, E.~Hill,
  K.~Krushelnick, A.~Maksimchuk, J.~Nees, S.J. Rose, A.G.R. Thomas, R.~Watt,
  J.C. Wood, V.~Yanovsky, and S.P.D. Mangles.
\newblock Demonstration of femtosecond broadband x-rays from laser wakefield
  acceleration as a source for pump-probe x-ray absorption studies.
\newblock \emph{High Energy Density Physics}, 35:\penalty0 100729, 2020.
\newblock ISSN 1574-1818.
\newblock \doi{https://doi.org/10.1016/j.hedp.2019.100729}.
\newblock URL
  \url{http://www.sciencedirect.com/science/article/pii/S1574181819300011}.

\bibitem[Faure et~al.(2018)Faure, Gustas, Guénot, Vernier, Böhle, Ouillé,
  Haessler, Lopez-Martens, and Lifschitz]{faure_review_2018}
J.~Faure, D.~Gustas, D.~Guénot, A.~Vernier, F.~Böhle, M.~Ouillé,
  S.~Haessler, R.~Lopez-Martens, and A.~Lifschitz.
\newblock A review of recent progress on laser-plasma acceleration at {kHz}
  repetition rate.
\newblock \emph{Plasma Phys. Control. Fusion}, 61\penalty0 (1):\penalty0
  014012, November 2018.
\newblock ISSN 0741-3335.
\newblock \doi{10.1088/1361-6587/aae047}.
\newblock URL \url{https://doi.org/10.1088\%2F1361-6587\%2Faae047}.

\bibitem[Salehi et~al.(2017)Salehi, Goers, Hine, Feder, Kuk, Miao, Woodbury,
  Kim, and Milchberg]{salehi_mev_2017}
F.~Salehi, A.~J. Goers, G.~A. Hine, L.~Feder, D.~Kuk, B.~Miao, D.~Woodbury,
  K.~Y. Kim, and H.~M. Milchberg.
\newblock {MeV} electron acceleration at 1 {kHz} with {\textless}10 {mJ} laser
  pulses.
\newblock \emph{Opt. Lett.}, 42\penalty0 (2):\penalty0 215, January 2017.
\newblock ISSN 0146-9592, 1539-4794.
\newblock \doi{10.1364/OL.42.000215}.
\newblock URL \url{https://www.osapublishing.org/abstract.cfm?URI=ol-42-2-215}.

\bibitem[Guénot et~al.(2017)Guénot, Gustas, Vernier, Beaurepaire, Böhle,
  Bocoum, Lozano, Jullien, Lopez-Martens, Lifschitz, and
  Faure]{guenot_relativistic_2017}
D.~Guénot, D.~Gustas, A.~Vernier, B.~Beaurepaire, F.~Böhle, M.~Bocoum,
  M.~Lozano, A.~Jullien, R.~Lopez-Martens, A.~Lifschitz, and J.~Faure.
\newblock Relativistic electron beams driven by {kHz} single-cycle light
  pulses.
\newblock \emph{Nature Photon}, 11\penalty0 (5):\penalty0 293--296, May 2017.
\newblock ISSN 1749-4893.
\newblock \doi{10.1038/nphoton.2017.46}.
\newblock URL \url{https://www.nature.com/articles/nphoton.2017.46}.

\bibitem[Ouillé et~al.(2020)Ouillé, Vernier, Böhle, Bocoum, Jullien, Lozano,
  Rousseau, Cheng, Gustas, Blumenstein, Simon, Haessler, Faure, Nagy, and
  Lopez-Martens]{ouille_relativistic-intensity_2020}
Marie Ouillé, Aline Vernier, Frederik Böhle, Maïmouna Bocoum, Aurélie
  Jullien, Magali Lozano, Jean-Philippe Rousseau, Zhao Cheng, Dominykas Gustas,
  Andreas Blumenstein, Peter Simon, Stefan Haessler, Jérôme Faure, Tamas
  Nagy, and Rodrigo Lopez-Martens.
\newblock Relativistic-intensity near-single-cycle light waveforms at {kHz}
  repetition rate.
\newblock \emph{Light Sci Appl}, 9\penalty0 (1):\penalty0 1--9, March 2020.
\newblock ISSN 2047-7538.
\newblock \doi{10.1038/s41377-020-0280-5}.
\newblock URL \url{https://www.nature.com/articles/s41377-020-0280-5}.
\newblock Number: 1 Publisher: Nature Publishing Group.

\bibitem[Baltuška et~al.(2003)Baltuška, Udem, Uiberacker, Hentschel,
  Goulielmakis, Gohle, Holzwarth, Yakovlev, Scrinzi, Hänsch, and
  Krausz]{baltuska_attosecond_2003}
A.~Baltuška, Th. Udem, M.~Uiberacker, M.~Hentschel, E.~Goulielmakis, Ch.
  Gohle, R.~Holzwarth, V.~S. Yakovlev, A.~Scrinzi, T.~W. Hänsch, and
  F.~Krausz.
\newblock Attosecond control of electronic processes by intense light fields.
\newblock \emph{Nature}, 421\penalty0 (6923):\penalty0 611--615, February 2003.
\newblock ISSN 1476-4687.
\newblock \doi{10.1038/nature01414}.
\newblock URL \url{https://doi.org/10.1038/nature01414}.

\bibitem[Ishii et~al.(2014)Ishii, Kaneshima, Kitano, Kanai, Watanabe, and
  Itatani]{ishii_carrier-envelope_2014}
Nobuhisa Ishii, Keisuke Kaneshima, Kenta Kitano, Teruto Kanai, Shuntaro
  Watanabe, and Jiro Itatani.
\newblock Carrier-envelope phase-dependent high harmonic generation in the
  water window using few-cycle infrared pulses.
\newblock \emph{Nature Communications}, 5\penalty0 (1):\penalty0 3331, February
  2014.
\newblock ISSN 2041-1723.
\newblock \doi{10.1038/ncomms4331}.
\newblock URL \url{https://doi.org/10.1038/ncomms4331}.

\bibitem[Nerush and Kostyukov(2009)]{nerush_carrier-envelope_2009}
E.~N. Nerush and I.~Yu. Kostyukov.
\newblock Carrier-{Envelope} {Phase} {Effects} in {Plasma}-{Based} {Electron}
  {Acceleration} with {Few}-{Cycle} {Laser} {Pulses}.
\newblock \emph{Phys. Rev. Lett.}, 103\penalty0 (3):\penalty0 035001, July
  2009.
\newblock \doi{10.1103/PhysRevLett.103.035001}.
\newblock URL \url{https://link.aps.org/doi/10.1103/PhysRevLett.103.035001}.

\bibitem[Zhidkov et~al.(2008)Zhidkov, Fujii, and Nemoto]{zhidkov_electron_2008}
Alexei Zhidkov, Takashi Fujii, and Koshichi Nemoto.
\newblock Electron self-injection during interaction of tightly focused
  few-cycle laser pulses with underdense plasma.
\newblock \emph{Phys. Rev. E}, 78\penalty0 (3):\penalty0 036406, September
  2008.
\newblock \doi{10.1103/PhysRevE.78.036406}.
\newblock URL \url{https://link.aps.org/doi/10.1103/PhysRevE.78.036406}.

\bibitem[Lifschitz and Malka(2012)]{lifschitz_optical_2012}
A.~F. Lifschitz and V.~Malka.
\newblock Optical phase effects in electron wakefield acceleration using
  few-cycle laser pulses.
\newblock \emph{New J. Phys.}, 14\penalty0 (5):\penalty0 053045, May 2012.
\newblock ISSN 1367-2630.
\newblock \doi{10.1088/1367-2630/14/5/053045}.
\newblock URL \url{https://doi.org/10.1088\%2F1367-2630\%2F14\%2F5\%2F053045}.

\bibitem[Budriūnas et~al.(2017)Budriūnas, Stanislauskas, Adamonis,
  Aleknavičius, Veitas, Gadonas, Balickas, Michailovas, and
  Varanavičius]{budriunas_53_2017}
Rimantas Budriūnas, Tomas Stanislauskas, Jonas Adamonis, Aidas Aleknavičius,
  Gediminas Veitas, Darius Gadonas, Stanislovas Balickas, Andrejus Michailovas,
  and Arūnas Varanavičius.
\newblock 53 {W} average power {CEP}-stabilized {OPCPA} system delivering 5.5
  {TW} few cycle pulses at 1 {kHz} repetition rate.
\newblock \emph{Opt. Express, OE}, 25\penalty0 (5):\penalty0 5797--5806, March
  2017.
\newblock ISSN 1094-4087.
\newblock \doi{10.1364/OE.25.005797}.
\newblock URL
  \url{https://www.osapublishing.org/oe/abstract.cfm?uri=oe-25-5-5797}.
\newblock Publisher: Optical Society of America.

\bibitem[Marcinkevičius et~al.(2001)Marcinkevičius, Juodkazis, Watanabe,
  Miwa, Matsuo, Misawa, and Nishii]{marcinkevicius_femtosecond_2001}
Andrius Marcinkevičius, Saulius Juodkazis, Mitsuru Watanabe, Masafumi Miwa,
  Shigeki Matsuo, Hiroaki Misawa, and Junji Nishii.
\newblock Femtosecond laser-assisted three-dimensional microfabrication in
  silica.
\newblock \emph{Opt. Lett., OL}, 26\penalty0 (5):\penalty0 277--279, March
  2001.
\newblock ISSN 1539-4794.
\newblock \doi{10.1364/OL.26.000277}.
\newblock URL
  \url{https://www.osapublishing.org/ol/abstract.cfm?uri=ol-26-5-277}.
\newblock Publisher: Optical Society of America.

\bibitem[Tomkus et~al.(2018)Tomkus, Girdauskas, Dudutis, Gečys, Stankevič,
  and Račiukaitis]{tomkus_high-density_2018}
Vidmantas Tomkus, Valdas Girdauskas, Juozas Dudutis, Paulius Gečys, Valdemar
  Stankevič, and Gediminas Račiukaitis.
\newblock High-density gas capillary nozzles manufactured by hybrid {3D} laser
  machining technique from fused silica.
\newblock \emph{Opt. Express, OE}, 26\penalty0 (21):\penalty0 27965--27977,
  October 2018.
\newblock ISSN 1094-4087.
\newblock \doi{10.1364/OE.26.027965}.
\newblock URL
  \url{https://www.osapublishing.org/oe/abstract.cfm?uri=oe-26-21-27965}.
\newblock Publisher: Optical Society of America.

\bibitem[Lehe et~al.(2016)Lehe, Kirchen, Andriyash, Godfrey, and
  Vay]{lehe_spectral_2016}
Rémi Lehe, Manuel Kirchen, Igor~A. Andriyash, Brendan~B. Godfrey, and Jean-Luc
  Vay.
\newblock A spectral, quasi-cylindrical and dispersion-free
  {Particle}-{In}-{Cell} algorithm.
\newblock \emph{Computer Physics Communications}, 203:\penalty0 66--82, June
  2016.
\newblock ISSN 0010-4655.
\newblock \doi{10.1016/j.cpc.2016.02.007}.
\newblock URL
  \url{http://www.sciencedirect.com/science/article/pii/S0010465516300224}.

\bibitem[Caron and Potvliege(1999)]{caron_free-space_1999}
C.~F.~R. Caron and R.~M. Potvliege.
\newblock Free-space propagation of ultrashort pulses: Space-time couplings in
  gaussian pulse beams.
\newblock \emph{Journal of Modern Optics}, 46\penalty0 (13):\penalty0
  1881--1891, 1999.
\newblock \doi{10.1080/09500349908231378}.
\newblock URL \url{https://doi.org/10.1080/09500349908231378}.

\bibitem[Beaurepaire et~al.(2014)Beaurepaire, Lifschitz, and
  Faure]{beaurepaire_electron_2014}
B.~Beaurepaire, A.~Lifschitz, and J.~Faure.
\newblock Electron acceleration in sub-relativistic wakefields driven by
  few-cycle laser pulses.
\newblock \emph{New J. Phys.}, 16\penalty0 (2):\penalty0 023023, February 2014.
\newblock ISSN 1367-2630.
\newblock \doi{10.1088/1367-2630/16/2/023023}.
\newblock URL \url{https://doi.org/10.1088\%2F1367-2630\%2F16\%2F2\%2F023023}.
\newblock Publisher: IOP Publishing.

\bibitem[Note1()]{Note1}
Note1.
\newblock In \protect \citep {nerush_carrier-envelope_2009} bubble asymmetry
  $\Lambda (l)$ is defined as the sum of transverse momenta of two electrons
  born with initial position $+l$ and $-l$ in the polarization plane.

\bibitem[Modena et~al.(1995)Modena, Najmudin, Dangor, Clayton, Marsh, Joshi,
  Malka, Darrow, Danson, Neely, and Walsh]{modena_electron_1995}
A.~Modena, Z.~Najmudin, A.~E. Dangor, C.~E. Clayton, K.~A. Marsh, C.~Joshi,
  V.~Malka, C.~B. Darrow, C.~Danson, D.~Neely, and F.~N. Walsh.
\newblock Electron acceleration from the breaking of relativistic plasma waves.
\newblock \emph{Nature}, 377:\penalty0 606--608, October 1995.
\newblock ISSN 0028-0836.
\newblock \doi{10.1038/377606a0}.
\newblock URL \url{http://adsabs.harvard.edu/abs/1995Natur.377..606M}.

\bibitem[Umstadter et~al.(1996)Umstadter, Chen, Maksimchuk, Mourou, and
  Wagner]{umstadter_nonlinear_1996}
D.~Umstadter, S.-Y. Chen, A.~Maksimchuk, G.~Mourou, and R.~Wagner.
\newblock Nonlinear {Optics} in {Relativistic} {Plasmas} and {Laser} {Wake}
  {Field} {Acceleration} of {Electrons}.
\newblock \emph{Science}, 273\penalty0 (5274):\penalty0 472--475, July 1996.
\newblock ISSN 0036-8075, 1095-9203.
\newblock \doi{10.1126/science.273.5274.472}.
\newblock URL \url{https://science.sciencemag.org/content/273/5274/472}.

\bibitem[Rousse et~al.(2004)Rousse, Phuoc, Shah, Pukhov, Lefebvre, Malka,
  Kiselev, Burgy, Rousseau, Umstadter, and Hulin]{rousse_production_2004}
Antoine Rousse, Kim~Ta Phuoc, Rahul Shah, Alexander Pukhov, Eric Lefebvre,
  Victor Malka, Sergey Kiselev, Fréderic Burgy, Jean-Philippe Rousseau, Donald
  Umstadter, and Daniéle Hulin.
\newblock Production of a {keV} {X}-{Ray} {Beam} from {Synchrotron} {Radiation}
  in {Relativistic} {Laser}-{Plasma} {Interaction}.
\newblock \emph{Phys. Rev. Lett.}, 93\penalty0 (13):\penalty0 135005, September
  2004.
\newblock \doi{10.1103/PhysRevLett.93.135005}.
\newblock URL \url{https://link.aps.org/doi/10.1103/PhysRevLett.93.135005}.
\newblock Publisher: American Physical Society.

\bibitem[Whittum et~al.(1991)Whittum, Sharp, Yu, Lampe, and
  Joyce]{whittum_electron-hose_1991}
David~H. Whittum, William~M. Sharp, Simon~S. Yu, Martin Lampe, and Glenn Joyce.
\newblock Electron-hose instability in the ion-focused regime.
\newblock \emph{Phys. Rev. Lett.}, 67\penalty0 (8):\penalty0 991--994, August
  1991.
\newblock \doi{10.1103/PhysRevLett.67.991}.
\newblock URL \url{https://link.aps.org/doi/10.1103/PhysRevLett.67.991}.
\newblock Publisher: American Physical Society.

\bibitem[Glinec et~al.(2008)Glinec, Faure, Lifschitz, Vieira, Fonseca, Silva,
  and Malka]{glinec_direct_2008}
Y.~Glinec, J.~Faure, A.~Lifschitz, J.~M. Vieira, R.~A. Fonseca, L.~O. Silva,
  and V.~Malka.
\newblock Direct observation of betatron oscillations in a laser-plasma
  electron accelerator.
\newblock \emph{EPL}, 81\penalty0 (6):\penalty0 64001, February 2008.
\newblock ISSN 0295-5075.
\newblock \doi{10.1209/0295-5075/81/64001}.
\newblock URL \url{https://doi.org/10.1209\%2F0295-5075\%2F81\%2F64001}.
\newblock Publisher: IOP Publishing.

\bibitem[Popp et~al.(2010)Popp, Vieira, Osterhoff, Major, H\"orlein, Fuchs,
  Weingartner, Rowlands-Rees, Marti, Fonseca, Martins, Silva, Hooker, Krausz,
  Gr\"uner, and Karsch]{popp_all-optical_2010}
A.~Popp, J.~Vieira, J.~Osterhoff, Zs. Major, R.~H\"orlein, M.~Fuchs,
  R.~Weingartner, T.~P. Rowlands-Rees, M.~Marti, R.~A. Fonseca, S.~F. Martins,
  L.~O. Silva, S.~M. Hooker, F.~Krausz, F.~Gr\"uner, and S.~Karsch.
\newblock All-optical steering of laser-wakefield-accelerated electron beams.
\newblock \emph{Phys. Rev. Lett.}, 105:\penalty0 215001, Nov 2010.
\newblock \doi{10.1103/PhysRevLett.105.215001}.
\newblock URL \url{https://link.aps.org/doi/10.1103/PhysRevLett.105.215001}.

\bibitem[Phuoc et~al.(2006)Phuoc, Corde, Shah, Albert, Fitour, Rousseau, Burgy,
  Mercier, and Rousse]{phuoc_imaging_2006}
Kim~Ta Phuoc, Sebastien Corde, Rahul Shah, Felicie Albert, Romuald Fitour,
  Jean-Philippe Rousseau, Fréderic Burgy, Brigitte Mercier, and Antoine
  Rousse.
\newblock Imaging {Electron} {Trajectories} in a {Laser}-{Wakefield} {Cavity}
  {Using} {Betatron} {X}-{Ray} {Radiation}.
\newblock \emph{Phys. Rev. Lett.}, 97\penalty0 (22):\penalty0 225002, November
  2006.
\newblock \doi{10.1103/PhysRevLett.97.225002}.
\newblock URL \url{https://link.aps.org/doi/10.1103/PhysRevLett.97.225002}.
\newblock Publisher: American Physical Society.

\bibitem[Bulanov et~al.(1998)Bulanov, Naumova, Pegoraro, and
  Sakai]{bulanov_particle_1998}
S.~Bulanov, N.~Naumova, F.~Pegoraro, and J.~Sakai.
\newblock Particle injection into the wave acceleration phase due to nonlinear
  wake wave breaking.
\newblock \emph{Phys. Rev. E}, 58\penalty0 (5):\penalty0 R5257--R5260, November
  1998.
\newblock \doi{10.1103/PhysRevE.58.R5257}.
\newblock URL \url{https://link.aps.org/doi/10.1103/PhysRevE.58.R5257}.
\newblock Publisher: American Physical Society.

\bibitem[Geddes et~al.(2008)Geddes, Nakamura, Plateau, Toth, Cormier-Michel,
  Esarey, Schroeder, Cary, and Leemans]{geddes_plasma-density-gradient_2008}
C.~G.~R. Geddes, K.~Nakamura, G.~R. Plateau, Cs. Toth, E.~Cormier-Michel,
  E.~Esarey, C.~B. Schroeder, J.~R. Cary, and W.~P. Leemans.
\newblock Plasma-{Density}-{Gradient} {Injection} of {Low}
  {Absolute}-{Momentum}-{Spread} {Electron} {Bunches}.
\newblock \emph{Phys. Rev. Lett.}, 100\penalty0 (21):\penalty0 215004, May
  2008.
\newblock \doi{10.1103/PhysRevLett.100.215004}.
\newblock URL \url{https://link.aps.org/doi/10.1103/PhysRevLett.100.215004}.
\newblock Publisher: American Physical Society.

\bibitem[Schmid et~al.(2010)Schmid, Buck, Sears, Mikhailova, Tautz, Herrmann,
  Geissler, Krausz, and Veisz]{schmid_density-transition_2010}
K.~Schmid, A.~Buck, C.~M.~S. Sears, J.~M. Mikhailova, R.~Tautz, D.~Herrmann,
  M.~Geissler, F.~Krausz, and L.~Veisz.
\newblock Density-transition based electron injector for laser driven wakefield
  accelerators.
\newblock \emph{Phys. Rev. ST Accel. Beams}, 13\penalty0 (9):\penalty0 091301,
  September 2010.
\newblock \doi{10.1103/PhysRevSTAB.13.091301}.
\newblock URL \url{https://link.aps.org/doi/10.1103/PhysRevSTAB.13.091301}.
\newblock Publisher: American Physical Society.

\bibitem[Beaurepaire et~al.(2015)Beaurepaire, Vernier, Bocoum, Böhle, Jullien,
  Rousseau, Lefrou, Douillet, Iaquaniello, Lopez-Martens, Lifschitz, and
  Faure]{beaurepaire_effect_2015}
B.~Beaurepaire, A.~Vernier, M.~Bocoum, F.~Böhle, A.~Jullien, J-P. Rousseau,
  T.~Lefrou, D.~Douillet, G.~Iaquaniello, R.~Lopez-Martens, A.~Lifschitz, and
  J.~Faure.
\newblock Effect of the {Laser} {Wave} {Front} in a {Laser}-{Plasma}
  {Accelerator}.
\newblock \emph{Phys. Rev. X}, 5\penalty0 (3):\penalty0 031012, July 2015.
\newblock \doi{10.1103/PhysRevX.5.031012}.
\newblock URL \url{https://link.aps.org/doi/10.1103/PhysRevX.5.031012}.

\bibitem[Corde et~al.(2013{\natexlab{a}})Corde, Thaury, Lifschitz, Lambert,
  Ta~Phuoc, Davoine, Lehe, Douillet, Rousse, and Malka]{corde_observation_2013}
S.~Corde, C.~Thaury, A.~Lifschitz, G.~Lambert, K.~Ta~Phuoc, X.~Davoine,
  R.~Lehe, D.~Douillet, A.~Rousse, and V.~Malka.
\newblock Observation of longitudinal and transverse self-injections in
  laser-plasma accelerators.
\newblock \emph{Nat Commun}, 4\penalty0 (1):\penalty0 1501, June
  2013{\natexlab{a}}.
\newblock ISSN 2041-1723.
\newblock \doi{10.1038/ncomms2528}.
\newblock URL \url{http://www.nature.com/articles/ncomms2528}.

\bibitem[Jackson(1999)]{jackson_classical_1999}
John~David Jackson.
\newblock \emph{Classical electrodynamics}.
\newblock Wiley, New York, {NY}, 3rd ed. edition, 1999.
\newblock ISBN 9780471309321.
\newblock URL \url{http://cdsweb.cern.ch/record/490457}.

\bibitem[Andriyash(2019)]{andriyash_synchrad}
Igor Andriyash.
\newblock Synchrad.
\newblock \url{https://github.com/hightower8083/synchrad}, 2019.

\bibitem[Corde et~al.(2013{\natexlab{b}})Corde, Ta~Phuoc, Lambert, Fitour,
  Malka, Rousse, Beck, and Lefebvre]{corde_femtosecond_2013}
S.~Corde, K.~Ta~Phuoc, G.~Lambert, R.~Fitour, V.~Malka, A.~Rousse, A.~Beck, and
  E.~Lefebvre.
\newblock Femtosecond x rays from laser-plasma accelerators.
\newblock \emph{Rev. Mod. Phys.}, 85\penalty0 (1):\penalty0 1--48, January
  2013{\natexlab{b}}.
\newblock ISSN 0034-6861, 1539-0756.
\newblock \doi{10.1103/RevModPhys.85.1}.
\newblock URL \url{https://link.aps.org/doi/10.1103/RevModPhys.85.1}.

\end{thebibliography}
\end{document}